\documentclass[3p, times, onecolumn]{elsarticle}

\usepackage{soul}
\usepackage{rotating}
\usepackage{graphicx}
\usepackage{amssymb}
\usepackage{blindtext}
\usepackage{makecell}
\usepackage{colortbl}
\usepackage{subcaption}
\usepackage{amsmath}
\usepackage[utf8]{inputenc}
\usepackage{csquotes}
\usepackage{cellspace}
\usepackage{array, tabularx, caption, boldline}
\usepackage{enumitem}
\usepackage{xcolor}
\usepackage{multirow}
\usepackage{booktabs}
\usepackage{bbding}
\usepackage{titlesec}
\usepackage{booktabs, caption}
\usepackage{listings}
\usepackage{algorithm} 
\usepackage{algpseudocode} 
\usepackage{multirow,tabularx}
\usepackage{parskip}
\usepackage{subcaption}
\usepackage{bbm}
\usepackage{hyperref}
\usepackage{todonotes}

\begin{document}

\begin{frontmatter}
\title{An Innovative Algorithm For Robust, Interactive, Piecewise-Linear Data Exploration}

\author[inst1]{Stephen Wright}
\affiliation[inst1]{organization={Department of Computer Science},
            addressline={University of York, Heslington}, 
            city={York},
            postcode={YO10 5GH}, 
            country={United Kingdom}}

\author[inst1]{Colin Paterson}

\begin{abstract}

Many mathematical modelling tasks (such as in Economics and Finance) are informed by data that is "found" rather than being the result of carefully designed experiments.  (e.g., tax returns, reported sales or confidence surveys, etc.)  This often results in data series that are short, noisy, multidimensional and contaminated with outliers, regime shifts, and confounding, uninformative or co-linear variables. 

In these circumstances it is critically important that the data exploration process is robust to these data issues and that it also allows interactive oversight by domain specialists who understand the underlying data generating processes. This paper is a working document that explains how the increasing availability of massively parallel processing allows the use of robust and intuitive techniques that have previously been under exploited due to their computational intensity. In particular, we conclude that a generalization of the Theil-Sen algorithm to reflect modes (rather than the median) in the parameter space distribution (of partial fits to the data) can provide a robust piecewise-linear fit to the data while also allowing for extensions to including elements of cluster analysis, regularization and cross-validation  in a unified (distribution free) approach that can:-

\begin{enumerate}
    \item Exploit piecewise linearity to reduce the need to pre-specify the form of the underlying data generating process.
    \item Detect non-homogeneity (e.g. regime shifts, multiple data generating processes etc.) in the data using an innovative non-parametric (Hamming-Distance/Affinity-Matrix) cluster analysis technique. 
    \item Enable dimension reduction and resistance to the effects of multi-co-linearity by including LASSO regularization as an integral part of the algorithm.  
    \item  Estimate measures of accuracy, such as standard errors, bias, and confidence intervals, without needing to rely on traditional distributional assumptions.      
\end{enumerate}

Taken together these extensions to the traditional Theil-Sen algorithm simplify the traditional process of parameter fitting by providing a single-stage analysis controlled by a multidimensional search of Scale/Parsimony/Precision hyper-parameters. This is intended to make the approach easily useable by the widest possible range of stakeholders.

These are early days in this research and the main limitation in this approach is that it assumes that compute power is infinite and compute time is small enough to allow interactive use. Both of these assumptions are only true for small datasets with limited number of explanatory variables. We expect these limitations to be significantly relaxed both with further work on the core algorithms and with the inevitable onward march of compute capability. To aid this process we have provided links to a browser based (Chrome) interactive demo app (with full source code) implemented in Python and TensorFlowJS.

\end{abstract}
\end{frontmatter}

\newpage
\section{Introduction}
 Economic and financial modelling is known to be particularly difficult for a number of data quality reasons \cite{griliches1984data} the majority of which can be traced back to the fact that most Economic data is “found” i.e. available as a by-product of some other activity. (Censuses, Tax collection, price and production surveys etc.) rather than being collected from well designed randomized experiments as you would do in the natural sciences. Hence it is important to involve people in the parameter estimation process that have a good qualitative grasp of the underlying data generating process (i.e. domain specialists such as economists) so they can oversee the quantitative analysis and so mitigate these problems.
 
  In order to enable this qualitative oversight of the modelling process, it is important to encapsulate as much of the boilerplate technicalities as possible (while still highlighting all the statistically important underlying trade-offs in the analysis). However this in turn requires that the encapsulated algorithms are as robust as possible to the above data quality issues. Hence our research into how the availability of massively parallel computing power might allow the exploitation of robust approaches that historically have been under utilised due to their high computational overhead.
  
  In particular, we have focused on the following:-
\begin{enumerate}
      \item \textbf{Non-parametric regression techniques} to efficiently extract parameters from small amounts of data while minimizing the impact of outliers. 
      \item \textbf{Piecewise linear parameter estimation} to avoid needing to choose a specific model form and allow for the possibility of regime shifts and/or confounding (categorical) explanatory variables while still giving easily explicable results.
      \item \textbf{An innovative Hamming Distance / Affinity Matrix approach to Cluster Analysis} to partition the input data into multiple piecewise linear fits.
      \item \textbf{Regularization and Cross validation} to mitigate the effect of multi co linearity and over fitting while also allowing the user to build more parsimonious models (dimension reduction).
       
\end{enumerate}

We argue in this paper that the ubiquitous availability of massively parallel computer power enables us to select a mutually coherent set of options for these issues which allows them to be viewed as a single algorithm (with hyper parameters). This avoids some traps for the unwary in the process of manually overseeing the parameter fitting process and removes the need to tailor these aspects of your model building pipeline to each new data set.

\begin{figure}
\centering
\includegraphics[width=0.8\textwidth]{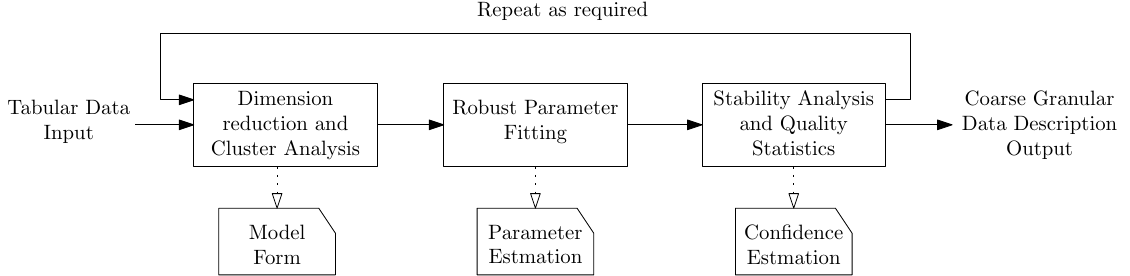}
\caption{Overview of the data exploration algorithm showing the three step process. These three steps can be repeated multiple times. The boxes under each process indicate the output from each step.}\label{fig:algorithm}
\end{figure}

  In particular, we conclude that a generalization of the Theil-Sen algorithm to reflect modes (rather than the median) in the parameter space distribution (of partial fits to the data) can provide a robust piecewise-linear fit to the data while also including regularization and cross-validation capabilities. Figure~\ref{fig:algorithm} illustrates the intended use of the algorithm. 

  In the next section (section 2) we give an introduction to piecewise linear parameter fitting and why we have chosen to select this as a standard form that can approximate a wide variety of data generating processes. This is followed by an overview of robust multivariate regression in section 3 to explain our selection of the Theil-Sen algorithm as the focus of our research. Section 4 then describes how the Theil-Sen algorithm can be extended to include regularisation, cross validation and an innovative Hamming Distance approach to cluster analysis. section 5 describes how we have evaluated the approach while section 6 draws conclusions and suggests further work.

  In the appendix we describe an interactive browser based demonstration app implementing this approach.
  
\section{Piecewise Data Fitting}
 Piecewise approximations to data are widely used in Mathematics and Engineering to solve problems that are otherwise intractable. For instance in mathematics it is useful for integration or optimisation tasks where some of the variables are non linear. ~\cite{rytin2004integration} \cite{schewe2012using} In engineering piecewise functions allow arbitrarily complex surfaces to be represented by lower order functions. These can be fixed or variable size as well as discontinuous, or continuous in the variable values and/or one or more of it's derivatives at the point where the segments overlap. \cite{schewe2012using} By adding more segments, we may use piecewise linear approximations to represent any non-linear function to any required accuracy.\cite{imamoto2008recursive}

There are many different ways to fit parameters to these piecewise functions such as the use of dummy variables in regression analysis with interactions (using dummy 0-1 variables for each segment)  \cite{warwicker2025efficient} or optimisation approaches given in standard python (and other) libraries such as \cite{jekel2019pwlf}. However these usually require you to specify the form of the underlying data generating function and any associated constraints. When exploring a new data set this information may not be easily available.

A similar approach which is of interest in the world of AI and Machine learning is regression or model trees as discussed in ~\cite{ilic2021explainable}. and ~\cite{raymaekers2024fast}. These learn an optimal splitting of the training data, as in a standard Decision Tree but the quality of the estimated values is improved  by fitting models (rather than the more usual constants) at each leaf.  ~\cite{cearliani2021linear}.

From this we can see that piecewise approximation of data sets is a well established and rich field of study in it's own right. For our purposes of economic and financial modelling, we are concerned with computer\textbf{ aided} modelling of processes that are fundamentally impossible to fully extrapolate from past data. ~\cite{king2020radical} This means that we need to merge quantitative statistical information as far as it can be reliably extrapolated into the future with theory and judgement taking over beyond that point. 

This in turn means that explicability of the model fitting process and ability to provide audit trails and manual oversight is fundamental to our work and so we have chosen to restrict ourselves to a simple subset of this field where we can apply robust discovery of multivariate piecewise linear approximations (without imposing any continuity, smoothness or uniqueness constraints) as discussed in the next section.

\section{Robust Multivariate Regression}

Various methods come under the heading of Robust Regression. These are all used to minimize the problem of outliers in the data however we wish to choose one method that is both as robust as possible and which also can be extended to deal with other aspects of the data such as the possibility of regime shifts etc. Therefore in this section we first consider the topic of robustness in regression analysis and then compare some of the regression algorithms at the more robust end of this spectrum in order to find the "best" for our purposes.

At a top level, the main difference between regression algorithms is explained by the different weightings applied to the residuals in the analysis.  e.g. in order of increasing robustness, these weighting schemes are:
\begin{itemize}
\item Sum of squared errors (Ordinary Least Squares or OLS) ~\cite{goldberger1964classical}
\item Sum of absolute deviations(Least Absolute Deviation or LAD) ~\cite{dodge2008least}
\item Sum of squares error on fitted points with absolute deviation on outliers (e.g. Huber loss function) or zero weight on outliers (Ransac)~\cite{andersen2008modern}
\item Median (rank) or integer count methods ~\cite{zuo2021general}
\end{itemize}
The Median (rank) or integer count weighting of residuals (e.g. concepts such as statistical depth) provides a very robust approach to parameter estimation and is non parametric in the sense of not assuming any particular distribution in order to fit a line to data. Unfortunately it is used much less often than other techniques because it is computationally intensive. As this is exactly the trade-off that is changing rapidly, this has been the main focus of our new work in this research phase. 

There are other regression methods referred to as non-parametric,  (often known as smoothing algorithms) which fit models to local data e.g. LOESS, Kernel methods, or Splines.  \cite{wasserman2006nonparametric}  However this extra freedom can come at the cost of being difficult to interpret and visualize the results. Therefore we have limited ourselves to a presumption of piecewise linearity and consider a discussion of these wider non parametric regression methods to be beyond the scope of this paper.

To understand our preferred class of distribution-free, linear-form regression models better, we must start with maximum likelihood ~\cite{rossi2018mathematical} This is where the likelihood of observing the given data values is calculated as a function of the parameters in a model. The best fit is the set of parameter values that generate the maximum likelihood of observing the given data values. This can be seen as searching parameter space for the maximum value of a function.

The integer count based regression techniques that we are considering can be posed (from a computational geometry perspective) ~\cite{rousseeuw2015statistical} as each data point generating a hyperplane in a maximum liklihood parameter space where cumulatively they construct their equivalent of a searchable likelihood function.

For example. in two dimensions, a line joining two data points corresponds to the vertex in parameter space where the two hyperplanes (generated by the two data points) intersect. Similarly, a straight line providing a perfect fit to "n" data points generates a set on "n" hyperplanes intersecting exactly at a single point in parameter space. 

An inexact fit results in these hyperplanes intersecting at "nearly" the same point in parameter space. This underlying process is easily extrapolated to higher dimensions even if visualisation becomes more difficult. The differences between the integer robust regression methods arises from how they capture the concept of "nearly intersecting".

The three main algorithms in this area (Regression Depth, the Hough Transform and Theil-Sen) focus on different aspects of this arrangement of hyperplanes Viz:

\textbf{Regression Depth} finds the region (one of the many formed between all the different hyperplanes in this arrangement) that is furthest (in the sense of the number of hyperplanes you need to cross) from an unbounded region in any direction. I.e. it can be thought of as the geometric median of that arrangement of hyperplanes (although the formal definition of multivariate medians is a topic of debate) ~\cite{rousseeuw1999regression}

The original \textbf{Hough Transform} is not strictly speaking a regression algorithm as it was developed in order to detect lines in images. We have included it in this comparison because it has strong similarities with regression problems, particularly when we wish to consider multiple fits to the data as we do in this paper. The original Hough Transform used a gradient/slope parametrisation. ~\cite{hart2009hough}  In this form it considers the parameter space to consist of an array of vote collectors, and when a hyperplane intersects with any collector, its vote is increased by one. Hence the best fit is defined as the parameter coordinates of the collector with the most votes. The transform has subsequently been further developed. We discuss the original form because this simplifies the comparison  with the other robust regression algorithms. The later generalised form is discussed in ~\cite{hart1972use}.  

\textbf{Theil-Sen} is a rank based variant of robust regression that is usually applied to two dimensional fitting problems. In this context it projects the position of vertices (hyperplane intersections) onto the slope axis and declares that the best-fit slope value is chosen to be the value taken by the median of this projected set of vertex positions~\cite{sen1968estimates} (back substitution is then used to derive the corresponding intersect value).

Essentially, the differences between the effectiveness of these three methods is related to how reliably it is possible to search the parameter space for the maximum value of their objective function. Considering these in turn:

\begin{itemize}
\item Regression Depth has a problem which is that if the data is very clean then all the hyperplanes will intersect at the same point (at the chosen search space resolution). if this happens then a parameter space search will not find any solution because the region(s) of interest are all approximately of size zero! Also, this approach does not handle cases where there may be multiple possible line fits to the data as it will tend to select a compromise solution half way between them.

\item The Hough Transform has a resolution problem. If too many vote accumulators are distributed evenly across parameter space, then the hyperplanes will not all intersect at the same accumulator. Rather they will form multiple local peaks in its vicinity.  This means that it is quite possible for an isolated local peak elsewhere in the accumulator array to have a count that is higher than any single one of the cluster of local peak counts in nearly adjacent accumulators at positions in parameter space that reflect what is arguably a better fit to the data. 

\item The Thiel-Sen Algorithm has the problem that it does not generalize well to multi-dimensions or (in it's standard form) work well with multi-modal data where its reliance on the median value can cause its results to become misleading and unstable. However, if we use modes in the projected vertex distributions rather than the median this opens up the possibility of generalizing it to apply to piecewise linear data fitting. Hence this algorithm has proved to be the most promising as discussed in the next section.
\end{itemize}

\section{Method - Extending The Theil-Sen-Siegel Algorithm}
\subsection{Overcoming a key limitation of Theil-Sen by using modes instead of the median }
The Theil-Sen algorithm is usually defined as a slope detector for two dimensional data sets. However for an "n" dimensional data set, it can be generalized to calculate \textbf{"candidate solutions"} (or vertices in parameter space) for every unique combination of "n" of the observed data points. In this generalization it then projects these vertices onto the appropriate parameter space axes and selects the median values of the resulting distributions along each axis as the "best fit" coordinates.

Unfortunately, if there is noise in the measurement of the explanatory variables (x directions) then candidate solutions that correspond to very steep positive slopes can sometimes (erroneously) be reported as having very steep negative slopes and vice versa. If this happens to a significant proportion of the candidate solutions this will result in a median value being selected which is highly unrepresentative of the desired best fit solution.

We have addressed this problem by using coordinates based on slope angle and not slope gradient. This results in a hyper-spherical coordinate system where very steep negative and positive slopes are now adjacent to each other. However this also introduces a circular distribution without defined end points. We therefore need to use the modes of this distribution rather than the median. 

The innovation of using modes rather than the median also impacts a second difficulty with the classic Theil-Sen algorithm which is that the median can be highly unstable if the distribution of projected candidate solution coordinates is in fact multi-modal. Conveniently addressing this problem by reflecting the possible multi mode characteristics in the data also opens up the possibility of fitting multiple local linear fits (one per mode) when the data is clearly multi-modal as discussed next. 

\subsection{Identifying modes in a noisy multi modal distribution}
Unfortunately detecting multiple modes in the data rather than the median introduces a new difficulty in its turn i.e. how to decide how many modes there might be and how to distinguish genuine modes from noise in the background data. Here we can exploit the same strategy as that used in Siegel’s median of medians ~\cite{siegel1982robust} version of Theil-Sen. 

In the Siegel algorithm the candidate solutions are partitioned into multiple sub sets (one for each of the "m" points in the input data). The candidate solutions in each such sub set all correspond to a fit through the one data point that they all have in common. The Siegel approach then finds the median coordinate values for each such subset of candidate solutions. The final "best fit" solution is deemed to be the median of these medians. 

In our variant of the Siegel approach, we partition the candidate solutions in the same way as the Siegel algorithm and then we assume that each data point can only be a member of one mode. This means we can search for the modes of a set of "m" uni modal distributions rather than an unknown number of modes in a multi modal distribution. We can then merge this large set of (often duplicate) uni modal modes into a smaller set of unique modes by employing an innovative Hamming Distance Affinity Matrix cluster algorithm in order to arrive at a single multi modal distribution with a known number of modes (piecewise linear fits to the data) as discussed in the next section.

\subsection{Choosing a unique subset of modes}
There are a wide variety of clustering algorithms that we could use to search for clusters of candidate solutions in parameter space, such as those based on centroids (e,g, K-Means), Those based on boundary markers (e.g. Support Vector clustering) Those based on density (e.g. DBSCAN or OPTICS) Those based on hierarchical clustering (e.g. Agglomeration or BIRCH) or Distribution-based clustering (e.g. Gaussian Mixture Model - GMM )

The discussion of the strengths and weaknesses of all the possible options is beyond the scope of this paper but it is worth noting that some such as K Means require a-priori knowledge of the number of clusters.  Others can be computational challenging, struggle with multi dimensional data or be sensitive to outliers in the data or to choice of initial parameters. A good overview of the principle methods in this area can be found in https://scikit-learn.org /stable/modules/clustering.html \cite{scikit-learn} Luckily, the unique design of the Theil-Sen algorithm opens up a new simple and robust possibility. 

Using a standard clustering algorithm we would need to search an "n" dimensional parameter space for clusters of candidate solutions. In the Theil-Sen algorithm we project the candidate solution coordinates onto the relevant axes then search for modes in these distributions of projected coordinates as discussed in previous sections. While this strategy can simplify the search for simple examples, for more complex ones some clusters in parameter space might generate overlapping modes in the axis distribution leaving us with a problem to disambiguate our modes. 

Luckily we know which candidate solutions vote for each mode in these distributions, we also know which data points were selected to generate each such candidate solution. Because there are many ("n" choose "m") candidate solutions with only a subset voting for each mode we can look at likelihood (vote count) of each data point in the underlying data being associated with a given mode (or piecewise linear fit). 

On it's own this does not help to disambiguate overlapping modes in the axis distributions. However if we look at pairwise probability, i.e. the likelihood of observing a given pair of data points together (in the same candidate solution) in each mode, then data points associated with the same underlying cluster (piecewise linear fit) will share a similar likelihood with all possible pairs of data points drawn from that cluster and zero with those from other clusters (or drawn from a mixture of clusters).

Repeating this process for all the modes found in the projected axis distributions will give us a pairwise likelihood of each data point being associated with the others. This is known as an affinity matrix. \cite{roffo2025origin} These are square symmetric matrices that are used to represent pairwise similarities between data points. If the data generating process consists of linear segments it can be sorted into block diagonal form where each block is a cluster (piecewise linear fit) to the underlying data.

There are many ways of assessing similarity of two points in parameter space  and most such measures exploit the concept of having a distance measure between the points such as:- 
\begin{itemize}
    \item Euclidean Distance between two points in parameter space 
    \item Manhattan Distance between two points in parameter space 
    \item Dot Product of the two vectors (vector direction and magnitude in parameter space)
    \item Angular (Cosine) separation of the two vectors in parameter space
    \item Hamming Distance (the number of positions at which two vectors of equal length differ)
\end{itemize}

A detail comparison of these alternative measures is beyond the scope of this paper, but more detail may be found in \cite{aggarwal2001surprising} \cite{wang2022dot} \cite{vom2024understanding} \cite{ref1}

We have chosen to use the pairwise likelihood of each data point being associated with each other data point in candidate solutions that voted for the mode of an axis wise distribution in order to construct an Affinity Matrix. This is because it arises naturally from the Theil-Sen algorithm. We then use a hamming distance measure to identify if any give pair of data points are likely to be in the same cluster. i.e. when comparing two rows of the affinity matrix with each other we count the number of columns where both vectors have a non zero element at that column position. 
\begin{enumerate}
    \item If this count is zero then the data points corresponding to those rows are definitely not in the same cluster.
    \item If this count is greater than half the count of the total number of non zero elements in either row then the two data points are deemed to be probably in the same cluster.
\end{enumerate}
This has the advantage that it provide an intuitive two dimensional approach to visualizing the number and size of clusters present in an "n" dimensional space so facilitates manual oversight and confidence building. Also it does not require the user to pre specify the number of clusters (but they may choose to provide a default which can guide a pruning process). However sorting the matrix into block diagonal form, identifying the number of such blocks and the data points associated with each is itself a challenge because the pairwise affinity is itself a noisy measure which may well give inconsistent and incomplete indications of cluster membership. This is a particular issue when the underlying data generating process does not actually consist of piecewise linear segments but is a continuous non linear function. Therefore we need to find the data point order that sorts  the affinity matrix into the "best" block diagonal form as discussed in the next section.

\subsection{Sorting The Affinity Matrix Into Block Diagonal Form}
For Data sets where there are expected to be a small number of clusters, and they are also expected to be quite distinct from each other, then the simplest way of ordering the data points is to use a parametric approach where they are ordered along an axis constructed from a linear combination of their coordinate values. However there is no easily identifiable "best" linear combination. For complex or noisy data sets this approach might rapidly become impractical. 

A theoretically attractive (if computationally more expensive option) is to construct a Minimum Spanning Tree. Because each data point has a path in the affinity matrix to each other data point (weighted by the affinity between them) we can find a set of paths that connects all the nodes in a tree together so that there is exactly one selected path through any one node. This is called a spanning tree. For any given graph there may be many possible spanning trees. If each path is weighted (e.g. by the similarity value between our data points) then one of these trees will exhibit the lowest sum of weights along it's set of constituent paths. This tree is a Minimum Spanning Tree (MST) \cite{mehlhorn2008minimum}\cite{jana2009efficient} 

There are multiple algorithms that can find these minimum scanning trees, and an evaluation has found that the best of these can be efficient \cite{gagolewski2024clustering} However, the computational overhead is still significant for larger problems and may be overkill as we are agnostic to the order of data points within each cluster.

Having ordered the data points using either of the above methods, they can be partitioned into sub-sets. This can be done by splitting the data point order where a (user-selected) minimum similarity value between two adjacent points is not met.

Given that the first approach above needs a search stage that may occasionally be unreliable for complex data sets and the second may be more elegant and robust but is always computationally expensive  we have explored a third  option (a heuristic algorithm) that merges the data point ordering and partitioning tasks.

Our Heuristic approach to identifying clusters in the affinity matrix starts by finding a data point that has the highest affinity matrix score (i.e. is in the largest cluster) and attaching to it a list of all the other data points that are clearly in the same block (see hamming distance calculation given above). These data points are then removed from the list and the process repeated recursively until there are no more data points to allocate.

For clean data this is all we need to do to find our clusters. For noisy data this may well result in multiple fits to sub sets of the data that are sub optimal. We therefore follow this first stage with a prune and merge stage based on recursively updating a cluster affinity matrix (based on the data point affinity of the cluster members). This allows us to identify and merge "similar" clusters using the same logic we have already used to identify "similar" data points. This recursive process ends when no further merges can be found.

Because we now have a set of clusters of data points, we can identify the best fit line for each such cluster as well as the residuals in data space between each data point and each best fit line. Hence we can identify a quality of fit measure for all the combinations of hyper parameter values (Scale/Parsimony/Precision) requested by the user. These can then be displayed as a heat map to aid user selection of the most appropriate hyper parameter values. 

Whichever way we choose to perform the above tasks, then for all the combinations of hyper parameter values  we now have both the data point order, and it is partitioned into clusters each with a best fit hyperplane. For each such cluster we can now convert our best fit hyperplane back from normalised to non normalised data space and report these as the best piecewise fit to the input data.

It is perhaps worth noting here that we are filtering the set of candidate solutions by a scale parameter (the maximum value of the normalised Cartesian distance between the associated data points). Because of our use of a pairwise affinity matrix this results in a rudimentary clustering (classification) capability (independent of the linearity displayed by members of the cluster). This may be observed in the Datasaurus "Dots" example dataset provided in our demonstration app. where the members of each dot are assigned to a separate cluster. Technically this is a classification result which can still be thought of as a piecewise linear fit but one where the additional linear fit information is  not relevant because it has a very wide confidence interval around the assigned slope values.

However this is not the end of the story because another potential advantage of basing our approach on generalising the Theil-Sen approach is that we can incorporate some other desirable properties of the data fitting process (such as Regularisation and elements of Cross validation)  as an intrinsic part of the algorithm. As discussed in the next section, this adds a further hyper parameter of regularisation penalty to the Scale and Precision values already calculated.

\subsection{Regularisation and Cross validation in the context of the Theil-Sen algorithm}

For an "n" dimensional data set with "m" data points, the Theil-Sen algorithm calculates candidate solutions in parameter space for every unique combination of "n" of the observed data points. i.e. this is an exact fit to those data points. It then selects the median value of all those candidate solutions as the "best fit" as discussed in previous sections.

In practical implementations of the Theil-Sen algorithm, such as that given in scikit-learn \cite{scikit-learn} the number of data points used to generate each candidate solution can be increased to "s" ("m"\textgreater"s"\textgreater"n"). When this is done then rather than an exact fit to each subset of points, an ordinary least squares fit is used to generate each candidate solution. A lower value of "s" leads to a higher breakdown point and a lower efficiency, while a higher value leads to a lower breakdown point and a higher efficiency. In the limit where "s"="m" the scikit-learn implementation of Theil-Sen is identical to ordinary least squares.

This use of inexact fitting to generate candidate solutions can be exploited to provide additional robustness properties. In particular if we replace the OLS fit with a LAD-LASSO (combined robust regression and regularisation) fit then we are making the candidate solutions less sensitive to outliers and multi-co-linearity or uninformative explanatory variables in the data. 

We can also recognise that by generating an inexact fit or candidate solution (based on a sub set of the data) then comparing it with other inexact fits based on different subsets of the data, we are in effect doing out-of-sample testing of the first fit which we are then repeating for all the other fits. The best fit overall is the one that agrees with "most" of it's peer group. 

We can see this as a rather extreme form of "Leave-p-out" cross-validation (LpOC) where "p" is a large proportion of the number of data points. This provides some degree of intrinsic ability to avoid over fitting as the Theil Sen algorithm by definition selects the candidate solution which agrees most with all it's out of sample equivalents as being the "best". For details on how different values of the size of the training set relative to the test set can affect the performance of the cross validation approach see \cite{nti2021performance}

Having identified our piecewise linear summary statistics for the input data by partitioning it into a number of clusters, we can calculate the quality of fit for each linear approximation to it's respective data point cluster in the usual way. 

The implementation of this unified algorithm is discussed in the following section.
\newpage
\subsection{Algorithm Overview}
The Algorithm described in this paper takes as input a data table where each row of the table is interpreted as a point in an "n" dimensional space where all but the last column is a continuous numeric explanatory (or independent) variable. The last column is a continuous dependant variable. The task is to identify the set of piecewise linear fits that are the best approximation to the relationship between the explanatory variables and the dependant variable.  Our algorithm to address this problem can be summarised in three stages viz:-

\begin{enumerate}
    \item Calculating a data point by data point affinity matrix as a function of Axis, Parsimony, Precision and Scale hyper-parameters. (see fig 2)
    \item For each such affinity matrix use a hamming distance metric to turn this into clusters of data points (i.e. possible piece wise linear fits)
    \item Use a quality of fit measure to select the "best" values of the hyper-parameters and then display these results.
\end{enumerate}

\vspace{1cm}
\textbf{function.... CalculateTheDataPointByDataPointAffinityMatrix(Axis,Parsimony,Precision,Scale)}
\begin{figure}[hbt!]
\includegraphics[scale=0.48]{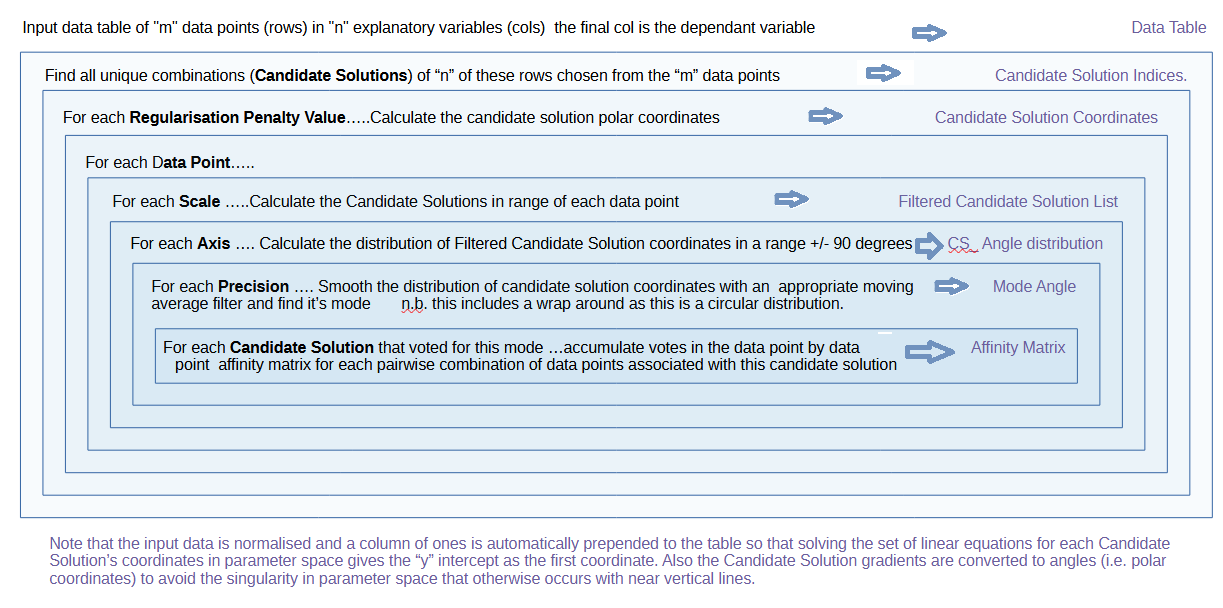}
\caption{The Affinity Matrix Calculation... }\label{fig:afPlot3}
\end{figure}

\textbf{return.... TheDataPointByDataPointAffinityMatrix}
\vspace{1cm}

Having calculated the affinity matrices (as a function of the hyper parameters) we have a measure of which data point pairs have been found together in the modes of the axis wise distribution of candidate solution coordinates. 

If the underlying data generating process is exactly piecewise linear then we should expect to see all combinations of the data points in each segment are flagged up in the affinity matrix. Hence we should then be able to sort the data point order such that the data points are in segment (or cluster) order and if we then sort the affinity matrix in this order we should see a perfect block diagonal form with one block per segment.

Of course the data generating process may be non-linear and contaminated with noise so it will not give an exactly block diagonal affinity matrix form. Hence we need to find the best approximation to this (as discussed in section 4.4). From a code perspective,  we have chosen to say that each data point has a signature which is the vector of it's association (or otherwise) with other data points in the affinity matrix. When the overlap of positive results between two data point signatures (i.e. the hamming distance between them) is greater than half the association count for either of those data points we mark them as likely to be in the same cluster. This gives us a simplified affinity matrix that we hope has removed the contamination caused by some candidate solutions occurring within an axis-wise mode by chance. 

We can now select the data point with the largest association count and use this as the seed member of the first (or current) cluster. Once we have added the associated data points to this cluster, we can remove the current cluster members from the set of active data points and recursively repeat this process until all the data points have been allocated an initial cluster (even if that is a cluster of size one).

Having generated an initial allocation of data points to clusters, we expect to see that for noisy data sets, we might get multiple clusters that are "similar" to each other so we need to consider pruning and merging such clusters. We can do this rather neatly by aggregating our data point affinity matrix to form a cluster affinity matrix and then generating a cluster hamming distance measure of similarity for these clusters as we had done previously for data point similarity. We can now recursively repeat this process to merge similar clusters until no further improvement is possible. 

Having now calculated our data point clusters as a function of Axis,Parsimony, Precision and Scale we can calculate the corresponding non-normalised fits to each cluster by doing our LAD-LASSO version of the Theil-Sen calculation on on each such sub set of data points. We can then calculate a quality of fit measure by summing the absolute value of the residual of each data point from its respective cluster (piecewise-linear) fit. We can then display the solution with the "best" combination of hyper parameters as an initial selection for subsequent manual oversight and data exploration.

Having arrived at our unified Robust Regression / Cluster Analysis /Regularisation algorithm, we now need to evaluate it's performance in practice as discussed in the next section.

\section{Evaluation}
In order to test the working of our algorithm, we have constructed a set of standard synthetic data sets and an associated demo app to run our algorithm. Running the app is discussed in Appendix A and the data sets are described below. They are split into two types. 

Firstly there are tests that can be done in two dimensions such as for robustness to outliers, non-linear data functions, regime shifts and confounding categorical variables. We have largely based these on two datasets provided in the literature (The Anscombe Quartet \cite{anscombe1973graphs} and the Datasaurus Dozen \cite{matejka2017datasaurus}).

These illustrate how data with identical summary statistics can have dramatically different underlying relationships. We have compared our piecewise linear fits to the results of applying some standard regression analysis routines to the Anscombe quartet. These comparison plots are shown in the appendix.

While this is not a fair comparison in the sense that the standard routines calculate a single fit where we know one of the quartet is more complex than that. What it does illustrate is that our piecewise linear fit prompts the user that there is some complexity that needs further investigation while routine unthinking use of standard regression analysis can provide highly misleading results (with no warning) in these circumstance.

We have also shown the results of applying our algorithm to the more complex Datasaurus Dozen data set. here we have not offered a comparison with standard regression routines because these are all intrinsic multi fit data examples. These  plots are also shown in the appendix.

To illustrate how to interpret the results in the appendix we have shown below a synthetic data set based on Simpsons paradox. This is a hypothetical exercise analysing the relationship between salary and years of experience with that firm for an industrial enterprise that has three entry grades, One for manual workers, another for graduate employees, and a third for high flyers poached from other companies late in their careers to fill senior management positions. However we have forgotten to separately flag this categorical variable.

Simply applying a standard regression algorithm will show that salary reduces with experience for this data set because the high paid employees have joined later in their careers and so have less experience with the firm.

Our algorithm (by default) suggests three fits to this hypothetical data (as seen in the Full scale high precision plot shown below. This is indicated as the best combination of hyper parameters in the quality of fit heat map. However that heat map suggests that if you are happy with a less precise fit then a small scale parameter will still give you three fits (indicating that for sub sets of the data, close to each other in problem space, salary increases with experience). However as the scale increases, to include more of the data, the quality of fit reduces but then increases again at full scale. This results in the full scale low precision plot that replicates the results you would get from a standard analysis and which depending on the purposes of your analysis you may prefer.

To complete the illustration, we have shown the data point by data point affinity matrix for the default fit. For a two dimensional task. This does not add much useful information in this case because it replicates the information we can eyeball from the scatter plot. However for multi dimensional tasks the affinity matrix will prove invaluable because it reports the quality of association between data points (and hence the number and quality of the underlying clusters) in two dimensions independent of the number of dimensions each cluster is spread over.

To be clear, we are not claiming here that our approach "solves" Simpsons paradox, it does not do this unless the confounding categorical variable in the analysis creates a sufficiently distinct set of clusters. What we are demonstrating here is that our presentation of a quality of fit heat map as a function of the hyper parameters in the analysis allows the manual identification and exploration of local minima as well as the usual global minima. Hence giving discretion to the user as to which is more relevant in their context. 

\begin{figure}[hbt!]
\includegraphics[scale=0.3]{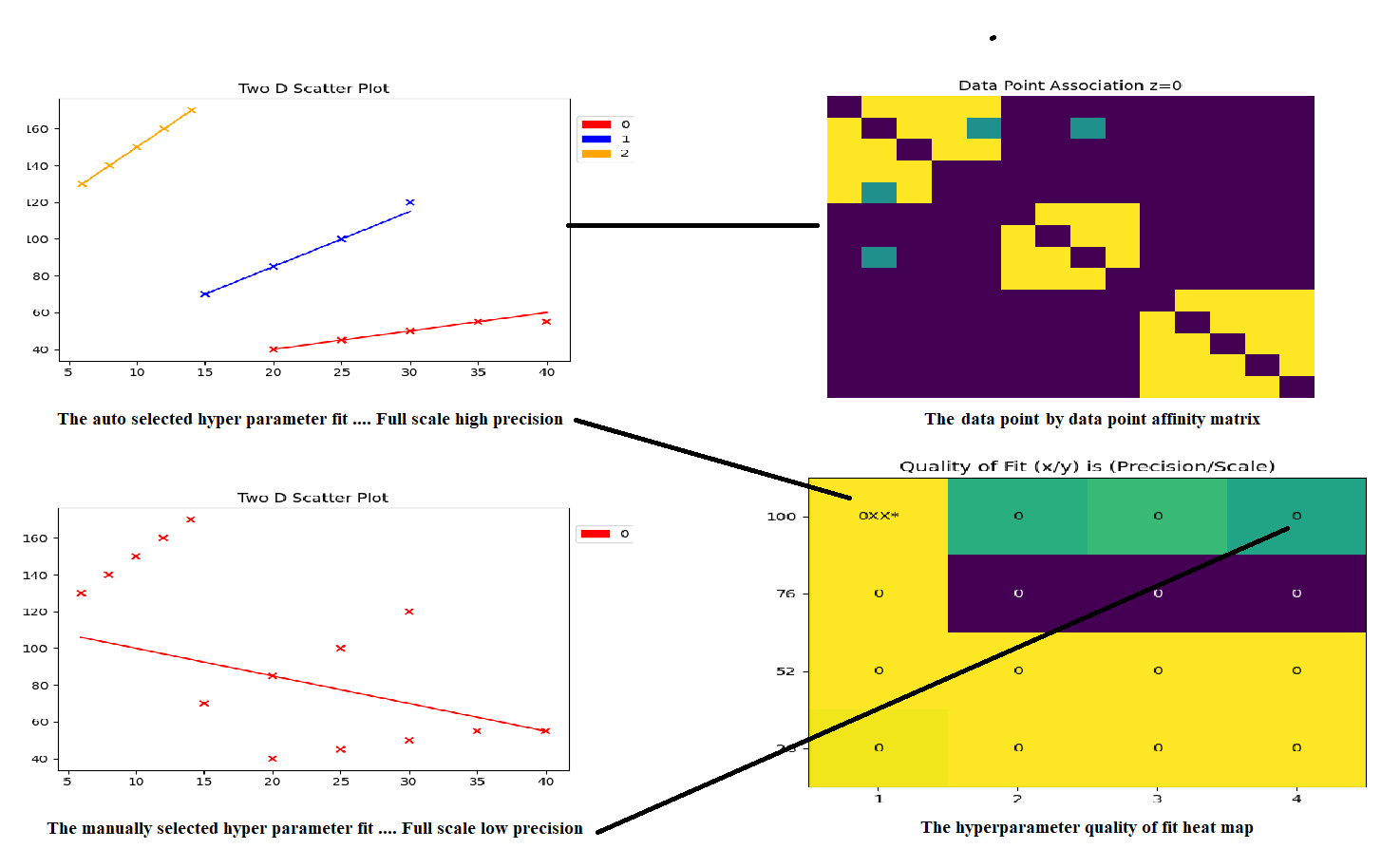}
\caption{Simpsons Paradox illustrates how to interpret the results of our algorithm}
\end{figure}

Finally, we have provided higher dimensional test data sets principally to test that the code is working when we generalise to multiple explanatory variables. These are not easy to illustrate. However because it is synthetic data it is trivial to confirm that the hyperplanes that are reported are the same as those we used to construct the test data. For the purposes of exposition in this paper we have not tried to illustrate the results of these test but leave this for the reader to confirm for themselves by running the canned examples (or their own data) using the supplied App.

The breakdown value is often quoted when comparing the robustness of alternative regression techniques \cite{morrison2021comparing}. This measure gives the smallest fraction of outliers in a data set that causes the regression to take on values that are essentially meaningless. It is used when fitting a single line to a set of data contaminated with noise. In our context, the noise data points might well be not just random contamination but another predictable linear relationship. 

In order to visualise this in a three dimensional example we have included a test data set (called "Clean Two Hyperplanes in 3D space") that consists of two 2D relationships embedded in different parts of a three dimensional space. For each piecewise linear fit half the points are relevant to the analysis but with an uninformative explanatory variable as well as the relevant explanatory variable. The other half of the data points are just noise and vice versa. If you view the results of the piecewise linear fit in each of the "x" axes you should see the scatter plot reflecting the fit to the relevant sub set of points for the relationship embedded in that dimension. 

The fact that the algorithm happily disentangles these two relationships is promising but We have not yet formalised how this might be characterised from the perspective of a breakdown value. However scikit-learn \cite{scikit} suggests that that conventional Theil-Sen algorithm "can tolerate arbitrary corrupted data (outliers) of up to 29.3 percent in the two-dimensional case.". We need to investigate how that transfers to the multi dimensional piecewise linear context.

\section{Conclusion and Suggested further work}
This paper describes some early work exploring how the ever increasing availability of massively parallel computing power might make data exploration more robust to mathematical modelling problems caused by data series that are short, noisy, multidimensional and contaminated with outliers, regime shifts, and confounding, uninformative or co-linear variables.

Traditional parameter estimation techniques usually make simplifying assumptions such as that the underlying data generating process has a known model form and is stationary with no outliers, regime shifts, uninformative or co-linear variables etc. While these assumptions may be appropriate for many parameter fitting tasks, when exploring a new data set they should not be automatically assumed to apply.  We have shown how the Theil-Sen robust regression algorithm can be extended to do piecewise linear fitting, include intrinsic LAD-LASSO regularisation and rudimentary classification in order to mitigate these data hygiene issues. 

We have produced a concept demonstrator that applies this algorithm to user supplied data tables. The core algorithms for this app are implemented in both Python and TensorFlowJS (see here \cite{wright2025simple} to view the code or run this App yourself) This demonstrator has been tested with a suite of known difficult synthetic data (see section 5 and the supplied App for details). 

The App applies a unified Robust Regression/Cluster Analysis/Regularisation to the input data. It has been tested with data series that are up to a few hundred points long with up to three explanatory variables. This assumes that you have a sufficiently capable GPU installed to run this code locally. You also need to be able to connect to the Pyodide web site. Larger data sets with more explanatory variables may result in long run times.

For clean data this algorithm works well. As the data becomes noisier, the number of fitted lines reported increases with more "similar" relationships being discovered. Also we start to observe some mathematically correct but in practice random alignment of points in the data. We are still calibrating these issues as well as working on prune and merge strategies for these fitted lines. We expect that a combination of automatic "similarity" tests and manual oversight will increase tolerance of this approach to these noise effects.

This work is motivated by the larger task of providing tools and techniques that deliver explicable parameter estimation in economic and financial modelling. As discussed this will often involve large numbers of small data sets. However, the issues addressed of merging cluster analysis, dimension reduction, robust parameter estimation and explicable results in a unified algorithm may well be of wider interest. 

As the research has progressed, it has become clear that there are significant parallels between this work and active areas of interest in machine learning.  In particular, the extended Theil-Sen algorithm that we use enables an innovative non-parametric Hamming Distance / Affinity Matrix approach to cluster analysis. Also the tasks of regression and classification lie at the heart of machine learning \cite{bishop2006pattern} It may therefore be interesting to look at generalising and scaling this work to be more relevant to this wider context as discussed in sub sections 6.2 and 6.3 below.

But first there are a number of questions raised by the current implementation that need further work which we describe in the next section. 

\subsection{Short term Technical and Human factor aspects}
Because our research is at an early stage we have focussed so far on correct operation of the code not performance issues. We now need to optimise the code for run time speed and memory usage and package it as a conventional python and Tensor Flow library to make it more easily available to other researchers in the field.

We need to do a complexity analysis to identify how rapidly the computation time varies with the number of dimensions and number of data points.

We need to Quantify the trade off between efficiency and resilience to over-fitting and/or co-linearity driven by increasing the sample size used to generate candidate solutions. \cite{nti2021performance}
    
We need to refine the cluster prune and merge heuristics to improve performance on noisy data. This needs to be informed by an intuitive feedback to the user on the under-fitting/ over-fitting trade off so they can interactively choose a "best" point on this spectrum.

In order to explore the sensitivity of the analysis to running on different explanatory variables we currently repeat it for each parameter space axis. However in  our experience to date, it turns out that even uninformative variables still generate a valid mode in the distribution of candidate solution coordinates (at zero) and if there are multiple potential fits in the data then on the uninformative variable axis these all overlap with each other at zero. 

Because we are using pairwise data point association in the affinity matrix our algorithm can still disambiguate these overlapping clusters at zero on this axis. Therefore if our cluster analysis works even for candidate solutions projected onto an uninformative variable, we may be able to either amalgamate the results from the separate axes into a single affinity matrix, or just do the analysis for an arbitrarily selected axis. Either way this will reduce the calculation time. This aspect needs to be more fully explored and the code refined as appropriate.

We need to exploit the cross validation (and/or bootstrapping) capabilities of our algorithm to calculate and display standard errors, bias, and confidence intervals around our parameter estimates.

The breakdown value is often quoted when comparing the robustness of alternative regression techniques \cite{morrison2021comparing}. This measure gives the smallest fraction of outliers in a data set that causes the regression to take on values that are essentially meaningless. scikit-learn \cite{scikit} suggests that that the conventional Theil-Sen algorithm "can tolerate arbitrary corrupted data (outliers) of up to 29.3 percent in the two-dimensional case.". We need to investigate how this measure transfers to the multi dimensional piecewise linear context.

We need to explore the capability of the integrated LASSO algorithm to suggest fitting models to aggregate data values in real world (hierarchical) economic data sets. for instance, economic activity is often reported at Global, Regional, Country, Sector, Sub-Sector and Enterprise levels of aggregation. In theory, regularisation should be able to identify when a more aggregate measure may be usefully exploited in a model as well as which sub categories have moved independently of that aggregate and so should perhaps be retained in their own right. This will need some human factors research if this capability is to be used as a standard feature of the analysis.

Finally, we should test the approach on a well researched real data set such as California Housing Data \cite{pace1997sparse}. Once the above work is done, there are some longer term objectives that suggest themselves as discussed below.

\subsection{Longer term Generalisation}
A (very) preliminary survey of the machine learning literature indicates that hallucination is a lively topic of interest. Our understanding is that data quality issues, are the initial "Garbage In" that eventually results in the plausible-sounding "Hallucination Out."

Most Machine learning libraries such as TensorFlow, PyTorch or SciKit-learn will use OLS as their default regression analysis because it's fast, has a closed-form solution, and the theory is well-developed. Other more robust approaches such as LAD and LASSO are also available as advanced options but need to be crafted into a bespoke "pipeline" customised to the circumstances of each project.

The underlying hope is that the dataset you are working with is in some sense homogeneous, the explanatory variables are relevant, the model form you are working with is well chosen and any limitations in the fitted model can be addressed by adding more data. In practice this is rarely the case because as George Box is quoted as saying "All models are wrong, but some models are useful"\cite{Box01121976}. 

This is why setting up the data often comprises a large proportion of the effort in a machine learning project \cite{lohr2014big}. It also explains why explicability (another hot button topic in machine learning) is important because deciding if a model is a sufficiently good approximation to be useful is an inherently qualitative decision requiring manual oversight.

Some of these data quality issues (such as missing values, inconsistent labelling, unrepresentative data etc.) will always need manual cleaning. However as we have seen from this research other aspects (such as outliers, unknown model form, data that is not homogeneous etc.) can be mitigated by exploiting a more robust fitting technique (at the expense of a higher computational workload). 

While a conventional machine learning project may address these issues by building a task specific pipeline with multiple assumptions introduced at each stage, we suspect that consolidating these into a single conceptual simple (piecewise linear fitting) stage can reduce the required data cleaning effort while simultaneously improving transparency and explicability.

In order to achieve this outcome, a much more detailed comparison of our parameter fitting approach with the standard machine learning frameworks will be needed. We will also need to  cater for categorical explanatory variables.  This can be achieved by  using a frequency approach \cite{hayfield2008nonparametric} that partitions the data into sub sets for each category. Or we could use one hot encoding  \cite{brownlee2020ordinal} that allocates a new explanatory variable to each category. 

In our context, the Data Partitioning approach to categorical explanatory variables is more attractive than the One Hot Encoding approach. This is because the former results in multiple smaller data sets so reducing our combinatorial explosion issue while the latter expands the number of explanatory variables hence aggravating the combinatorial explosion that is the key limiting factor on the practicality of our algorithm as discussed in the next section, 

\subsection{Longer term Scalability}
Our initial research has focussed on short noisy data sets with a small number of explanatory variables and a small number of piecewise linear fits with real time interaction with the user. In the wider context of Machine learning, this interactive aspect is less relevant as researchers in this field will often think of elapsed time in the training phase of days not minutes however if this work is to be relevant in this wider context then it is important that it proves to be scalable to larger more complex problems in time constraints acceptable to that community.

This scalability issue comes in three forms, number of data points, number of dimensions, and complexity of the resulting set of parameter fits. Each of these can be addressed in order to widen the applicability of this approach.

Randomisation, Stratified Sampling and/or Multi Scale Analysis can all address the combinatorial explosion that occurs with the number of data points. However devising detailed algorithms (consistent with the rest of the approach) and exploring the trade off between the different options will need considerable work.

The combinatorial explosion that occurs in our approach with the number of dimensions is aggravated by our use of a brute force solution to solving the LAD-LASSO fitting process. For small numbers of dimensions this is a very simple algorithm that allows us to get an exact solution to this regularisation task in a small fixed number of steps. This is ideal for running thousands of such calculations in parallel. 

As the number of dimensions increases, the number of vertices that need testing in the underlying LP optimisation task suffers a combinatorial explosion that compounds the one simultaneously occurring caused by the number of data points. Hence for larger numbers of dimensions a coordinate descent algorithm will start to give better results as discussed in our companion paper \cite{wright2025simple}. The trade off between increased complexity of the algorithm, run time speed and the extent to which the upper limit on the practical number of dimensions can be increased needs to be evaluated.

Finally, as the number of piecewise linear fits increases, it will become ever more difficult to interpret the results. If the paradigm could be expanded to describe this set in the form of a decision (or model) tree this would simplify the interrogation of the results by the user. This is also a promising line of research being developed by others (i.e. specifically into model trees \cite{raymaekers2024fast}). The difference is that in this latter approach, the model trees are built recursively from the initial partition to the final leaves. In our case we have a set of leaves that we want to organise into some hierarchical tree. Working out how to do this will require some research. However if it is possible to include a model tree stage in our paradigm then this may also allow us to improve the current rudimentary classification capabilities in our approach. This would be a major benefit in out quest for a simplified and unified standard module that (computation power allowing) could be widely applied in diverse parameter fitting tasks.

\newpage
\appendix
\section{Running the interactive (Chrome) browser based demonstration app}
In the comments section associated with this paper on ArXiv we provide a link to a (Chrome) browser based app demonstrating this algorithm.

When you click on this link it will load a simple minimalist (static HTML) IDE. When the red "Python Loading..." message disappears, you can hit the "Run Mode" button at the top left and then the demo app will start up. You may wish to open the Chrome debug window, by pressing "Ctrl+Shift+J" in order to monitor progress and flag up any un-anticipated problems. When the "Compiling" and then the red "Loading Packages...." message at the top right of the screen is finished then the You can start a new analysis by pasting in a table of data. This should be in standard Excel format with column names in  the top row and numeric dependant variable in the final column (all the other columns are deemed to be explanatory variables). Alternatively, you can select and copy a pre-canned example from the "Test Data" tabs and paste this into the input data table.

You can choose how many steps there are in each dimension of the hyper parameter search space by setting the precision, scale and regularisation option selectors. You can then hit the "Calculate" button. As this calculation is done locally, the speed will depend on the size of the data set you have entered and the capability of your GPU. However it is likely that your browser will report an unresponsive page (perhaps multiple times). If this happens please be patient and wait for the calculation to complete. 

Please note that the calculations have only been tested for up to 100 data points and three explanatory variables.

When the calculation is complete, the results tab will be displayed. This has four fields viz:.....
\begin{enumerate}
    \item A quality of fit heat map for the selected hyper parameter search space.
    \item A scatter plot showing the selected fit(s) to the data points if there is only one explanatory variable. Otherwise, the scatter plot shows the residuals in the user selected axis.
    \item An affinity matrix showing the strength of association between data points
    \item A results table listing the piecewise linear fits that have been calculated.
    \item A console log displaying miscellaneous associated data such as cluster membership vectors etc.
\end{enumerate}

The run will automatically choose one combination of hyper parameter values as the "best" and display the corresponding affinity matrix and scatter plot. The quality of fit heat map will mark the selected "best" option with "XXX". Note that you will often wish to vary the initial selection of hyper parameter values in order to exploit your domain specific knowledge of the data set in question. To override the auto selected hyper parameter values, make your alternative  selections in the option boxes at the top of the screen and press the "Redisplay" button.

This demonstration app has implemented the algorithm in both Python and TensorflowJS versions. the former in order to make the code accessible to a wide variety of users and the latter to exploit the massively parallel processing capabilities if you have a high end local GPU. To view the code that ran these calculations hit the "Design Mode" button at the top left. Then select the "Code" tab that appears at the top left. Choose a code module from the Code Tree that this displays and then press the "Edit" button at the bottom of this code tree display. A code editor will then pop up and you can view all the relevant code.

For more comprehensive instructions on how to use this IDE to explore the demo app, hit the "Help" button at the top left of the screen.

Please note that due to vectorisation and availability of maths routines in TensorFlowJS and Python the two implementations give slightly different results. We are currently working to reconcile these differences. We are also continuing to explore trade-offs between speed and robustness so the online code base may be updated from time to time to reflect this. Also because the TensorflowJS code can have compatibility problems with integrated GPUs and really requires a powerful GPU we have disabled this version by default in the demo app.(however the relevant TensorflowJS code can still be viewed)

This browser based version of our demo app is simply offered as a zero installation, zero configuration, minimal learning curve, existence proof that the algorithm works for small tasks (if you have the patience to wait for it). It also allows you to interrogate all the source code behind the app. It is not expected to provide a practical working environment for real development work. As a matter of practicality, we do our development work using the desktop version of this IDE (see https://github.com/Steve--W/XComponents) on a PC equipped with a high end Nvidia GPU (a GTX 4060 or above).
 \newpage
\section{ - Selected 2D test data sets}

This appendix  illustrates the performance of our algorithm on some selected test data sets. 

Statistician Frank Anscombe devised the demonstration (shown in fig 3) in 1973 to illustrate the limitations of basic summary statistics for describing realistic datasets. \cite{anscombe1973graphs} It uses four datasets, each with 11 (x,y) points: The Sum, Average and St.dev of both X and Y are identical in all four charts.

\begin{figure}[hbt!]
\includegraphics[scale=0.6]{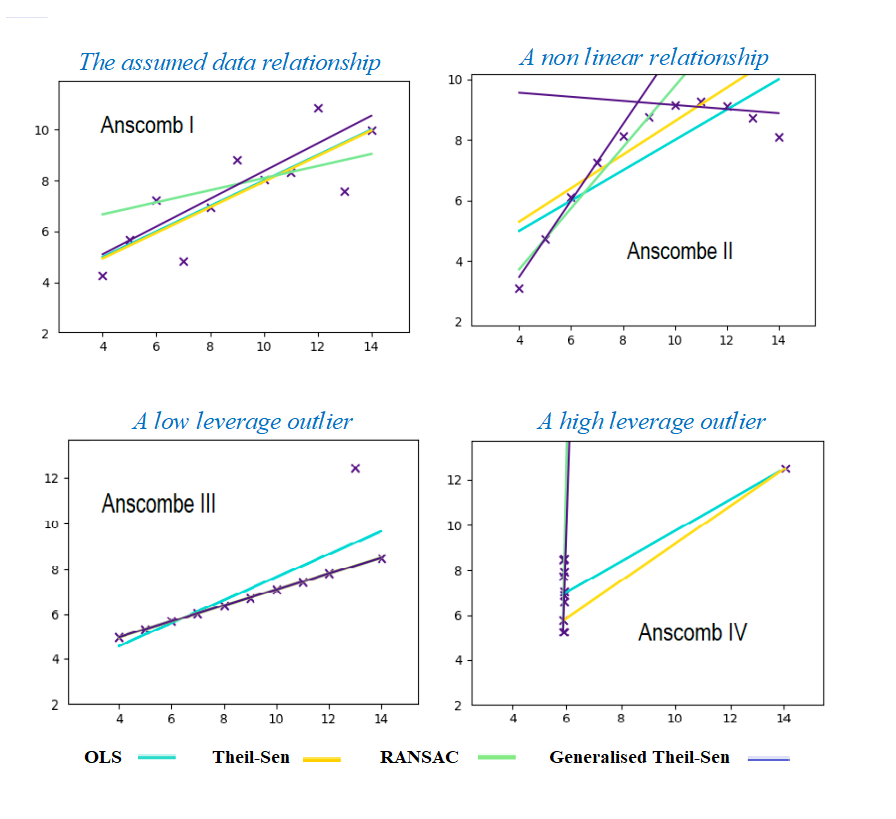}
\caption{The Anscombe Quartet... }\label{fig:afPlot3b}
\end{figure}

\begin{table}[hbt!]
\centering
    \caption {Summary statistics shared by all four datasets\label{tab:AnscombeSummary}}
    \begin{tabular}{lcl} \toprule
    \textbf{Property} & \textbf{Value} & \textbf{Accuracy} \\ \midrule
    Mean of x & 9 & exact\\
    Sample variance of x & 11 & exact\\
    Mean of y & 7.50 & to 2 decimal places\\
    Sample variance of y & 4.125 & $\pm$0.003\\
    Correlation between x and y & 0.816 & to 3 decimal places\\
    Linear regression line & $y = 3.00 + 0.500x$ & to 2 and 3 decimal places respectively\\
    Coefficient of determination & 0.67 & to 2 decimal places\\
    \bottomrule
     \end{tabular}
\end{table}
\newpage
The following pages show charts for non linear, Confounding (categorical) and regime shift synthetic data examples for both clean and noisy data sets.

\begin{figure} [hbt!]
\includegraphics[scale=0.5]{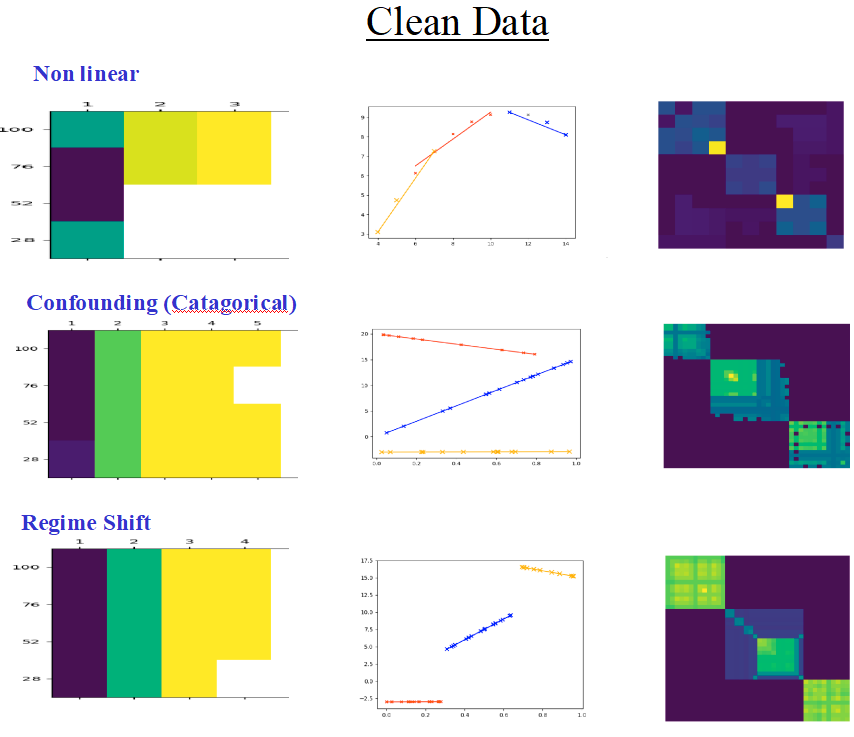}
\caption{Clean Data... }\label{fig:afPlot4}
\end{figure}

\newpage

\begin{figure}[hbt!]
\includegraphics[scale=0.5]{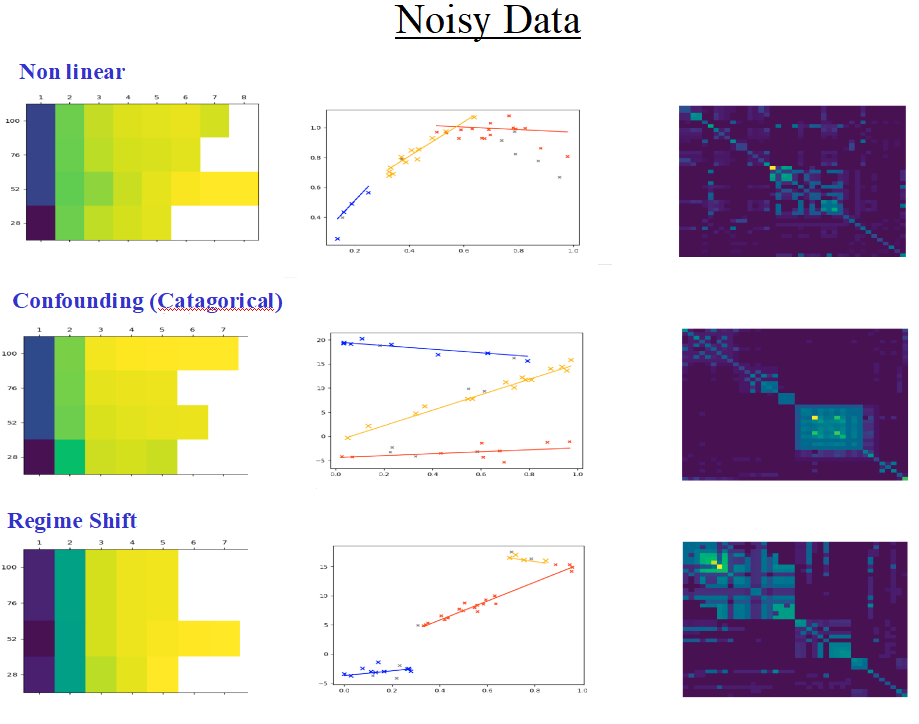}
\caption{Noisy Data... }\label{fig:afPlot5}
\end{figure}

The next page shows some Synthetic data sets selected from the Datasaurus dozen. These have nearly identical descriptive statistics to two decimal places, yet have very different different form when graphed. \cite{matejka2017datasaurus} 

\newpage These are not expected to be representative of the types of relationships being modelled in practice, but are presented here because they are widely recognised in the literature as a more modern example of how to illustrate the limitations of basic summary statistics for describing realistic data sets. Hence they are a good test of the ability of our algorithm to deal with more than usually difficult data sets.

\begin{figure}[hbt!]
\includegraphics[scale=0.5]{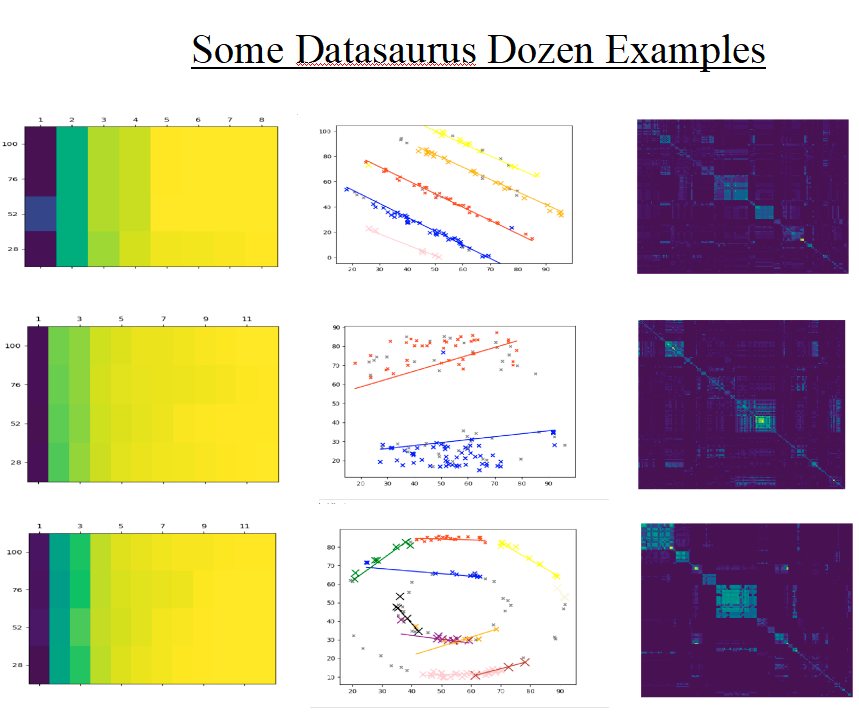}
\caption{Datasaurus 1... }\label{fig:afPlot6}
\end{figure}

\begin{figure}[hbt!]
\includegraphics[scale=0.5]{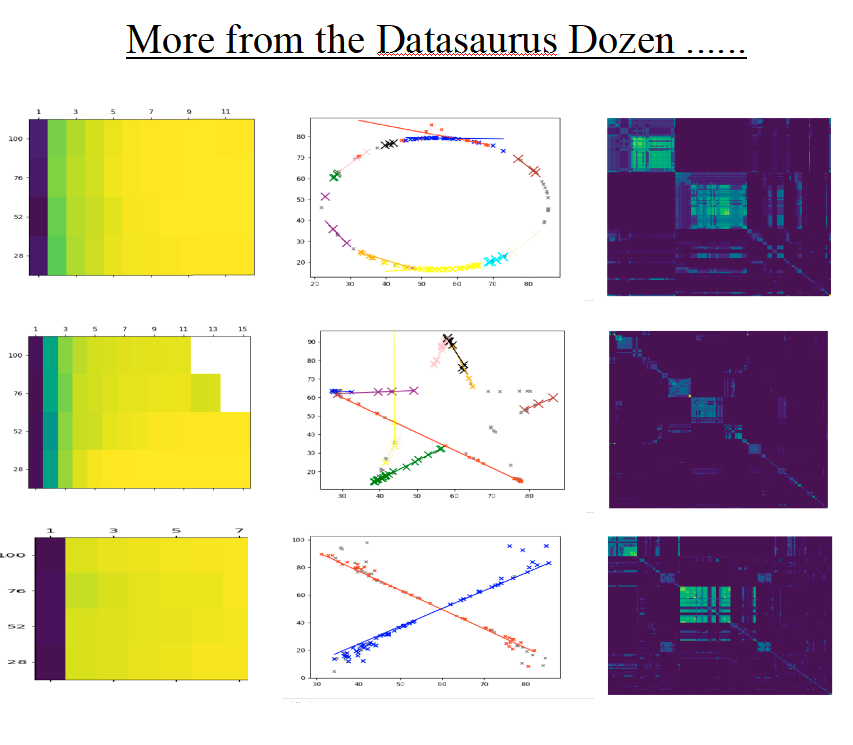}
\caption{Datasaurus 2... }\label{fig:afPlot7}
\end{figure}
\newpage

\bibliographystyle{ieeetr}
\bibliography{refs}
\end{document}